\documentclass[%
reprint,
superscriptaddress,
amsmath,amssymb,
aps,prl
]{revtex4-1}

\usepackage{graphicx}
\usepackage{dcolumn}
\usepackage{bm}

\usepackage[mathscr]{euscript}

\newcommand{\ud}{\mathrm{d}}
\newcommand{\bu}{\mathbf{u}}
\newcommand{\bp}{\mathbf{p}}
\newcommand{\bA}{\mathbf{A}}
\newcommand{\R}{\mathbb{R}}
\newcommand{\lP}{\langle{\bf P}|}
\newcommand{\rA}{|{\bf A}\rangle}

\newcommand{\pbar}{\bar{p}}

\usepackage{color}
\definecolor{dkgreen}{rgb}{0,.6,0}
\definecolor{purple}{rgb}{.75,0,1}
\definecolor{orange}{rgb}{1,.4,0}

\usepackage{placeins}

\begin{document}
	
\preprint{APS/123-QED}

\title{Simultaneous measurement of multiple parameters of a subwavelength structure\\ based on the weak value formalism}

\author{Anthony Vella}
	\affiliation{The Institute of Optics, University of Rochester, Rochester, NY 14627, USA}
\author{Stephen T.~Head}
	\affiliation{The Institute of Optics, University of Rochester, Rochester, NY 14627, USA}
\author{Thomas G.~Brown}
	\affiliation{The Institute of Optics, University of Rochester, Rochester, NY 14627, USA}
\author{Miguel A.~Alonso}%
	\email{alonso@optics.rochester.edu}
	\affiliation{The Institute of Optics, University of Rochester, Rochester, NY 14627, USA}
	\affiliation{Aix-Marseille Univ., CNRS, Centrale Marseille, Institut Fresnel, \\UMR 7249, 13397 Marseille Cedex 20, France}

\date{\today}

\begin{abstract}
	A mathematical extension of the weak value formalism to the simultaneous measurement of multiple parameters is presented in the context of an optical focused vector beam scatterometry experiment. In this example, preselection and postselection are achieved via spatially-varying polarization control, which can be tailored to optimize the sensitivity to parameter variations. Initial experiments for the two-parameter case demonstrate that this method can be used to measure physical parameters with resolutions at least 1000 times smaller than the wavelength of illumination.
\end{abstract}

\pacs{Valid PACS appear here}
\maketitle


The concepts of weak value and weak measurement were introduced by Aharonov, Albert and Vaidman in 1988 \cite{Aharonov_1988,Tamir_2013,Svensson_2013} as an alternative to the standard measurement formalism of quantum mechanics. For a quantity associated with an operator $B$, a standard measurement is related to the {\it expected value} $\langle\Phi|B|\Phi\rangle/\langle\Phi|\Phi\rangle$, where $\Phi$ is the state vector for the quantum state being measured. Since the state is normalized, the inner product in the denominator is typically taken as unity. Clearly, for Hermitian operators, this expected value is real and limited to the range of values spanned by the eigenvalues of $B$. On the other hand, weak measurements are based on {\it weak values} defined as $\langle\Phi_{\rm post}|B|\Phi_{\rm pre}\rangle/\langle\Phi_{\rm post}|\Phi_{\rm pre}\rangle$, where $\Phi_{\rm pre}$ and $\Phi_{\rm post}$ are preselected and postselected states. It is easy to see that there is no bound to a weak value since the denominator can be made arbitrarily small by appropriate preselection and postselection. In fact, weak values need not even be real-valued.
Weak measurements have been employed, for example, to measure very small angular deviations with great precision \cite{Hosten_2008,Dixon_2009,Dennis_2012,Merano_2014}.

While weak values are usually presented in the language of quantum theory, their formalism applies to classical measurements as well. Indeed, not only can the essential elements of some of their most successful experimental applications be explained classically \cite{Hosten_2008,Dixon_2009} (with some exceptions \cite{Hallaji_2016}), but many pre-existing important interferometric techniques can be interpreted in terms of weak values, in which the preselected state describes the illumination, and postselection is achieved by a filtering process of the resulting light, either spatially, directionally, temporally, spectrally, or in polarization. Three examples of this are phase contrast microscopy \cite{Zernike_1942_1,Zernike_1942_2}, which earned Zernike the Nobel Prize in Physics in 1956, differential interference contrast microscopy \cite{Murphy_2001}, and off-null ellipsometry \cite{Pedersen_1986}.

\begin{figure}
	\centering
	\includegraphics[width=\linewidth]{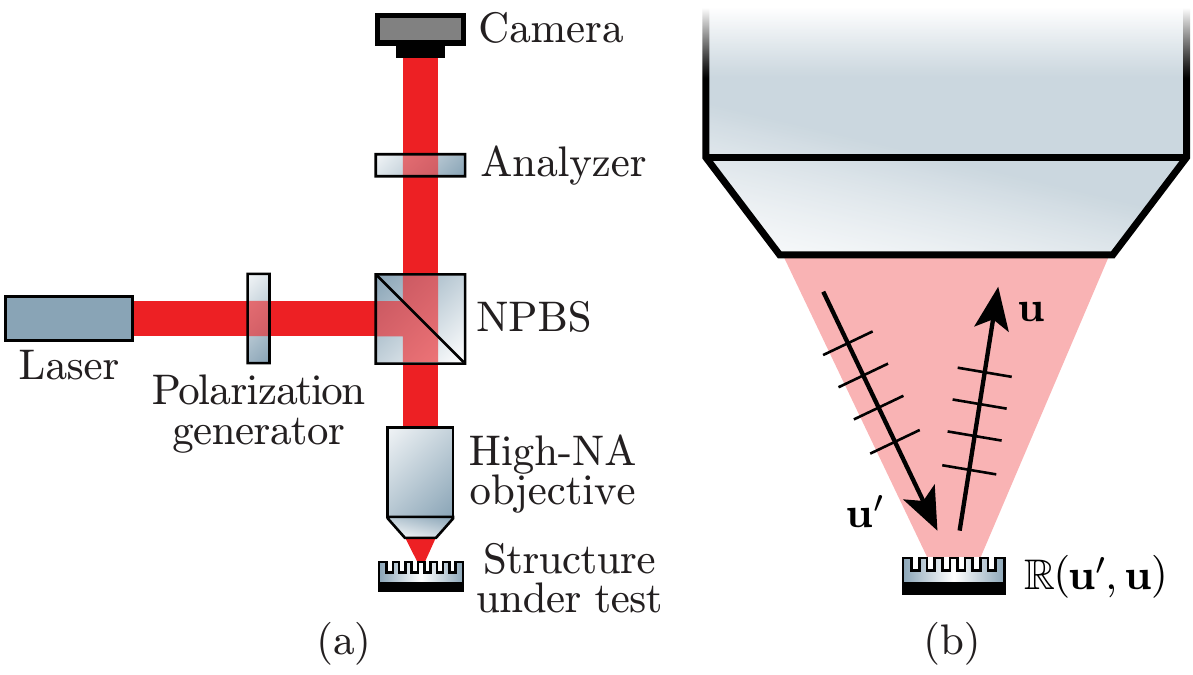}
	\caption{(a) Schematic of a focused vector beam scatterometry experiment in which preselection and postselection are achieved via spatially-varying polarization control. NPBS = non-polarizing beamsplitter. (b) The structure's scattering matrix $\R(\bu',\bu)$ provides the coupling between incident and reflected plane waves with directions $\bu'$ and $\bu$.}
	\label{fig:FBS_schematic}
\end{figure}

In this Letter, we present a method for simultaneous measurement of multiple parameters, inspired by the weak value formalism. We concentrate on the measurement of several morphological parameters of a periodic structure with subwavelength features. This example has practical applications in the semiconductor industry, in which manufacturers require precise measurements of integrated circuit components (e.g., stacked silicon wafers), often on sub-nanometer scales. Typical parameters of interest include the period, critical dimension (CD), overlay error, line edge roughness, trench depth and profile,  film thickness, and wafer alignment and orientation \cite{Diebold_2001,ITRS_2015,DenBoef_2016}. Here we present the theory for the measurement of several parameters, as well as experimental results for a two-parameter measurement of CD and orientation angle, which must be controlled during the etching process to reduce overlay error. The measurement scheme, which we refer to as \emph{focused vector beam scatterometry}, is illustrated in Fig.~\ref{fig:FBS_schematic}(a). A polarized beam is focused onto the structure, then the scattered light is re-collimated, passed through a polarization analyzer, and measured. The key of this method is to design the incident polarization and the analyzer (either of which may be spatially inhomogeneous in general) to optimize the sensitivity of the measurement. The experimental layout is similar to that of coherent Fourier scatterometry \cite{Boher_2004,Kumar_2014}, which also uses a focused beam but lacks optimized polarization control.

Let us begin with a mathematical description of this approach. In the linear regime, the test structure is characterized by its scattering matrix $\R(\bu',\bu)$, which provides the coupling between an incident plane wave and a reflected plane wave with directions specified by the transverse direction cosines $\bu' =(u_x',u_y')$ and $\bu =(u_x,u_y)$, respectively (see Fig.~\ref{fig:FBS_schematic}(b)). The directional variables $\bu'$ and $\bu$ are mapped by the objective lens onto the spatial pupil positions of the collimated input and output beams. After focusing, the incident field is represented by a vector angular spectrum, $\bA(\bu')$, which gives the complex amplitude and polarization of the plane wave component in the direction $\bu'$. The angular spectrum of the field reflected by the structure is then given by
\begin{equation}
\bA_{\rm R}(\bu )=\int\R(\bu',\bu)\bA(\bu ')\,\ud^2u'.
\end{equation}
After collection and collimation by the lens, the analyzer transmits a given polarization ${\bf P}(\bu)$ at each point, and the transmitted intensity is measured at the CCD. This measured intensity can be written as
\begin{equation}
I(\bu)=|{\bf P}^\dagger(\bu)\bA_{\rm R}(\bu)|^2=|\lP\R\rA|^2,
\end{equation}
where ${\bf P}^\dagger$ is a conjugate transpose and
\begin{equation}
\lP\R\rA(\bu)=\int{\bf P}^\dagger(\bu)\R(\bu',\bu)\bA(\bu')\,\ud^2u'.
\end{equation}

The goal is to simultaneously measure a set of $N$ morphological parameters of the structure, denoted as $\bp=(p_1, p_2,...,p_N)$, such as those mentioned above. Because we are considering small ranges of the values of interest, we can assume that the scattering matrix has approximately linear dependence on these parameters:
\begin{equation}
\R(\bu',\bu;{\bf p})\approx\R_0(\bu',\bu)+\sum_{n}p_n\R_n(\bu',\bu),
\end{equation}
where the index of summation runs from 1 to $N$. For simplicity, these parameters are normalized to be dimensionless and for their ranges of variation of interest to correspond to $|p_n|\le1$, with $p_n=0$ corresponding to the nominal structure. The measured intensity is then
\begin{subequations}
\begin{align}
I(\bu;\bp)&\approx\left|\lP\R_0\rA+\sum_{n}p_n\lP\R_n\rA\right|^2\\&=|\lP\R_0\rA|^2\left|1+\sum_{n}p_n\frac{\lP\R_n\rA}{\lP\R_0\rA}\right|^2.\label{eq:I_weak}
\end{align}
\end{subequations}
The form shown in Eq.~(\ref{eq:I_weak}) is factorized to emphasize the connection to weak measurements, where $\lP\R_n\rA/\lP\R_0\rA$ is analogous to the weak value of $\R_n$. (More precisely, it is the weak value of $\R_0^{-\alpha}\R_n\R_0^{-1+\alpha}$ with preselected and postselected states $\R_0^{1-\alpha}\rA$ and $\lP\R_0^{ \alpha}$, respectively, for any real $\alpha$.)

The key to a good measurement is to tailor $\bA$ and ${\bf P}$ so that these weak values are real and have variations on the order of unity, and so that their dependences on $\bu$ are as distinguishable as possible. This is achieved by letting
\begin{equation}
\lP\R_0\rA=-\sum_{n}\pbar_n(\bu)\lP\R_n\rA,
\label{condition}
\end{equation}
where $\pbar_1(\bu),\ldots,\pbar_N(\bu)$ are a set of real functions of $\bu$. The output intensity distribution is then given by
\begin{equation}
I(\bu;\bp)\approx\left|\sum_{n}[p_n-\pbar_n(\bu)]\lP\R_n\rA(\bu)\right|^2.\label{eq:I(u;p)_pbar}
\end{equation}
The design of a good measurement system then reduces to a suitable choice of pupil-dependent reference functions $\pbar_n(\bu)$, from which $\bA(\bu)$ and/or ${\bf P}(\bu)$ can be determined. In our setup we choose the analyzer to be uniform (${\bf P}$ is constant), implying that a spatially-varying input polarization $\bA(\bu)$ would be required for optimal sensitivity. Let us consider the simple case where the test structure introduces no directional coupling, i.e., $\R(\bu',\bu;\bp)=\R(\bu;\bp)\,\delta(\bu -\bu ')$. This is the case, for example, for a uniform multilayer thin film or a periodic structure with subwavelength period. The condition in Eq.~(\ref{condition}) can be written as
\begin{equation}
{\bf P}^\dagger\!\left(\R_0(\bu)+\sum_{n}\pbar_n(\bu)\R_n(\bu)\right)\!\bA(\bu)
=0,
\end{equation}
which leads to the following solution for the incident angular spectrum:
\begin{equation}
\bA(\bu)=
A(\bu)
\left[\!\begin{array}{cc}0&1\\-1&0\end{array}\!\right]\!
\left(\R_0+\sum_{n}\pbar_n\R_n\right)^{\!\!T}
{\bf P}^*,\label{eq:A(u)_solution}
\end{equation}
where $A(\bu)$ is an arbitrary envelope function, the superscript $T$ denotes a matrix transpose, and ${\bf P}^*$ is a complex conjugate. For structures that couple different directions, the solution for $\bA$ in terms of ${\bf P}$ and the functions $\pbar_n$ is similar although more complicated, sometimes requiring an iterative process. 

\begin{figure*}
	\centering
	\includegraphics[width=\linewidth]{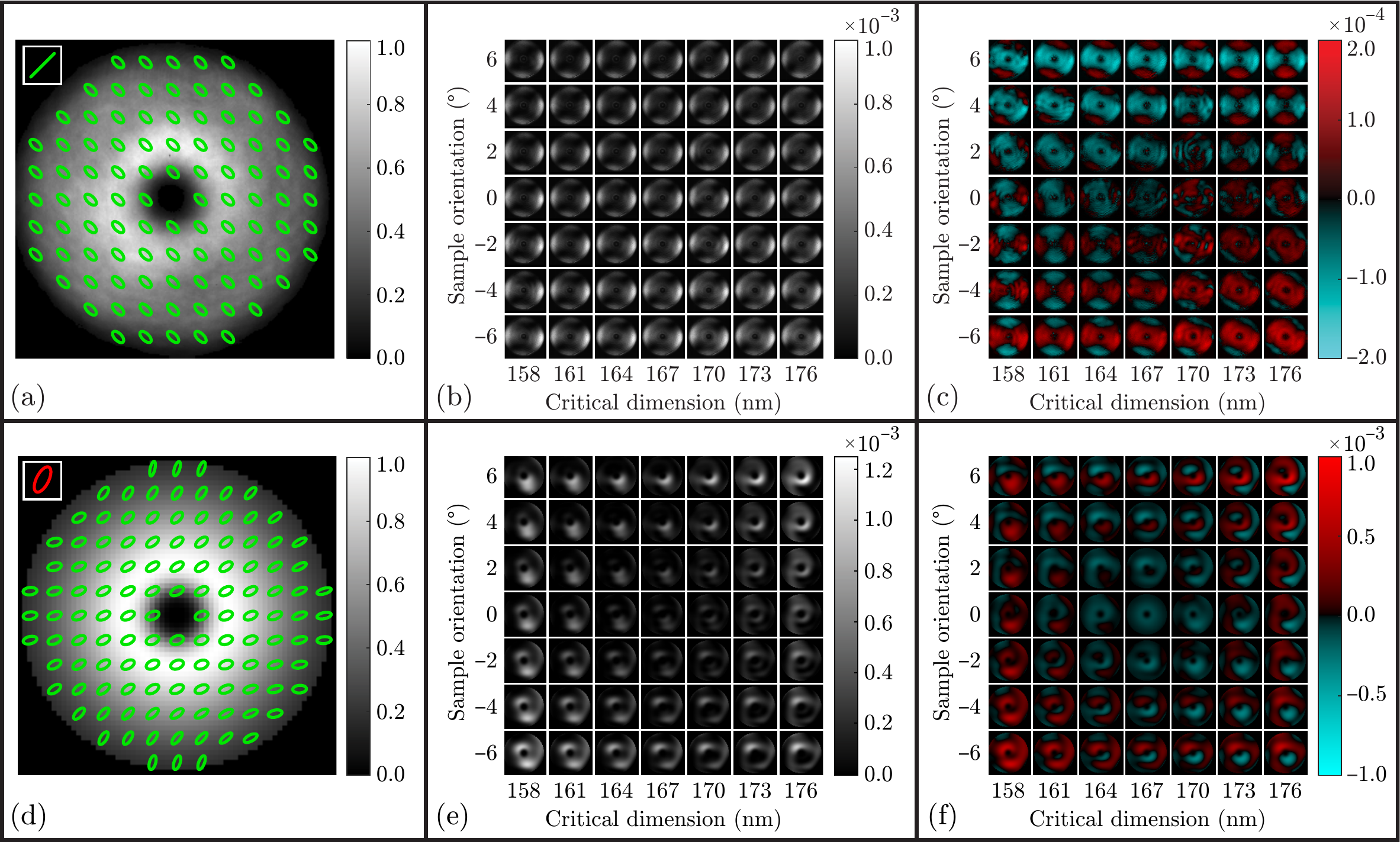}
	\caption{Experimentally measured (top row) and optimized simulated (bottom row) input polarizations and output intensity distributions for a two-parameter measurement of critical dimension and sample orientation. (a,d) Incident polarization and normalized intensity after transmission through the beamsplitter. An annular apodization profile was used to prevent unwanted backreflections on-axis. The transmitted polarization ${\bf P}$ of the output analyzer is shown in the upper left corner. Right- and left-handed polarization states are represented by green and red ellipses, respectively. (b,e) Output intensity distributions for 49 experimental/simulated measurements, normalized to the same scale as the input intensity. The axis labels indicate the parameter values associated with the intensity distributions shown in each row and column. (c,f) Differences between each intensity distribution and the mean distribution over the parameter range.}
	\label{fig:in_out_intensities}
\end{figure*}

Note that by substituting Eq.~(\ref{eq:A(u)_solution}) into Eq.~(\ref{eq:I(u;p)_pbar}), the measured intensity can be written as a multivariate quadratic function of the form
\begin{equation}
I(\bu;\bp)\approx|A(\bu)|^2\sum_{n',n''}(p_{n'}-{\bar p}_{n'})\Gamma_{n'n''}(p_{n''}-{\bar p}_{n''}), \label{eq:Iquad}
\end{equation}
where the coefficients $\Gamma_{n'n''}(\bu)$ (which depend on $\R_n$ and ${\bf P}$) are the elements of a real, positive semidefinite Hermitian $N\times N$ matrix. In practice, for a fixed input polarization state, one can expand Eq.~(\ref{eq:Iquad}) and calculate the quadratic coefficients directly from a set of experimental calibration images of reference structures with known parameters. The advantage of this approach is that it accounts for some sources of systematic error, including any deviation between the experimentally achieved input polarization and the theoretical distribution. Using the calibrated intensity profile, the physical parameters associated with an observed intensity from an unknown structure may then be determined using maximum likelihood estimation (MLE) techniques. The estimation uncertainty is inversely proportional to the square root of the eigenvalues of the Fisher information matrix, which can be computed from $I(\bu;\bp)$. For further details on the use of MLE in this context, see Ref.~\cite{Vella_2018_MLE_arxiv}.

As an example of this method, we now present the results of a two-parameter measurement of a one-dimensional lamellar silicon grating structure with 0.4 $\mu$m period. The two measured parameters were the grating's critical dimension (CD) and its orientation angle (relative to horizontal) in the plane perpendicular to the optical axis. The illumination wavelength was 1.064 $\mu$m, so the subwavelength grating diffracted only a single propagating order, introducing no directional coupling. The basic layout for the experiment is contained in Fig.~\ref{fig:FBS_schematic}; additional details on the experimental apparatus and implementation can be found in Appendix I.

The preliminary measurements presented here were taken using a uniform linear analyzer oriented at 45$^\circ$ and a uniform incident polarization. The input polarization, illustrated in Fig.~\ref{fig:in_out_intensities}(a), was chosen to minimize the transmission through the analyzer, resulting in the closest possible approximation of the conditions for optimal sensitivity to parameter variations. Future measurements are planned using a spatially-varying polarization generator (currently under development) and a uniform elliptical analyzer. Fig.~\ref{fig:in_out_intensities}(d) shows a simulation of the optimal input polarization and analyzer for this configuration, which were designed to maximize the eigenvalues of the Fisher information matrix over the parameter range of interest. The optimized functions $\pbar_1(\bu)$ and $\pbar_2(\bu)$ associated with this input polarization are provided in Appendix II.

A total of 49 measurements, shown in Figs.~\ref{fig:in_out_intensities}(b,e) for the experimental and simulated cases, were collected for seven structures with critical dimensions between 158 nm and 176 nm oriented at angles between $-6^\circ$ and $6^\circ$. The variations in intensity over this parameter range can be visualized by subtracting the mean intensity from each measurement, as seen in Figs.~\ref{fig:in_out_intensities}(c,f). Notice that the maximum variation of the experimental intensity from the mean is approximately 20\% as large as the peak intensity. In comparison, the simulated spatially-varying polarization produces intensity variations up to 70\% of the peak value, making the effects of the structure parameters more easily distinguishable.

\begin{figure}
	\centering
	\includegraphics[width=.95\linewidth]{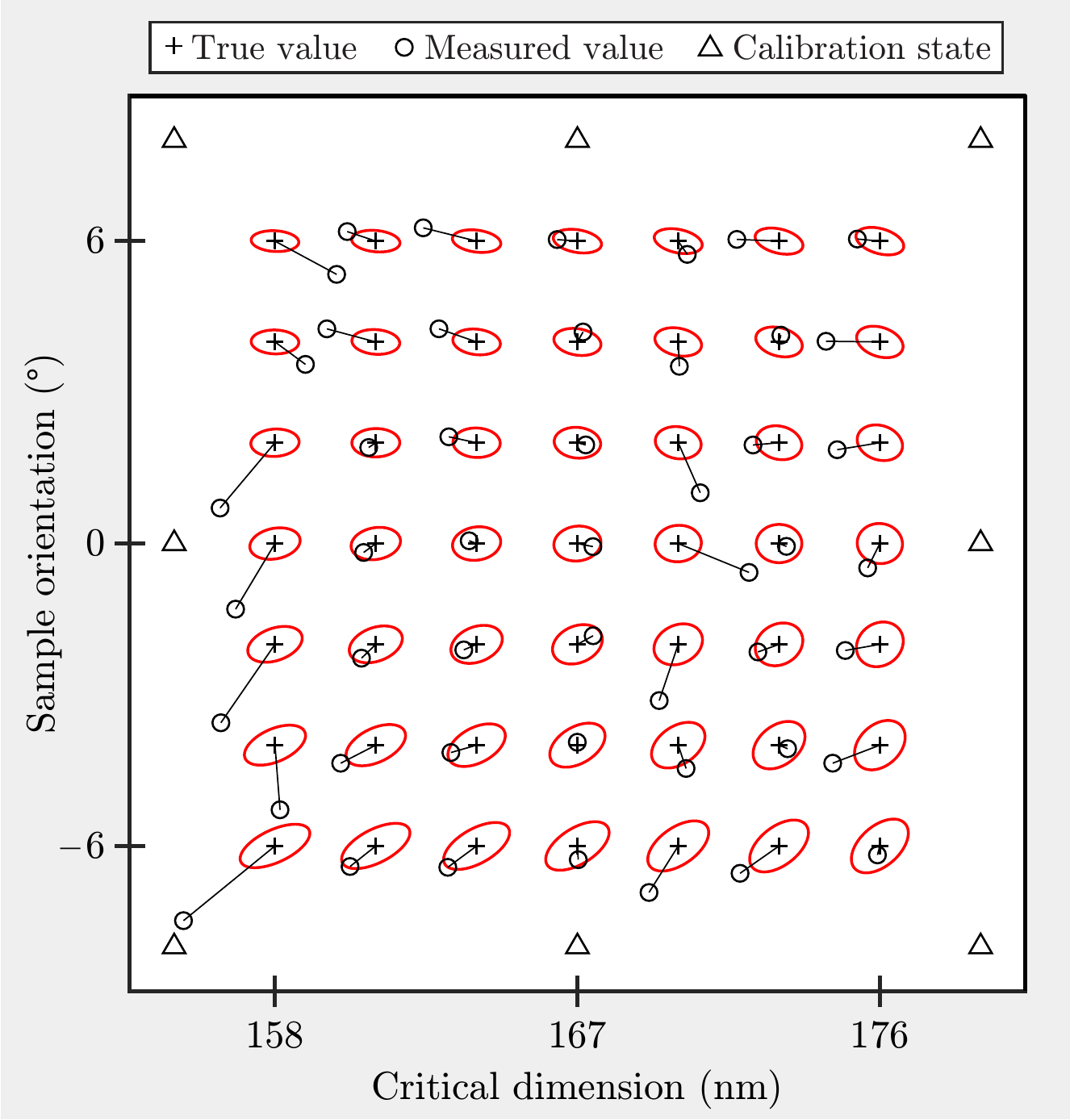}
	\caption{Estimated parameters from each of the 49 images shown in Fig.~\ref{fig:in_out_intensities}(b). Error bars connect each estimate to the associated true parameter values. The ellipses represent the minimum standard deviation error expected from a measurement of 7500 photons.}
	\label{fig:param_retr}
\end{figure}

The parameters associated with each experimental image were estimated using MLE techniques and compared to the ``true'' parameter values obtained from a series of focused ion beam (FIB) measurements and manual readings of the sample's rotation stage. The uncertainties in these assumed ``true'' values may be as large as 1 to 2 nm and $0.2^\circ$, respectively. The true and measured parameters for each measurement are plotted in Fig.~\ref{fig:param_retr}, along with the parameter values associated with eight additional reference measurements that were used for calibration purposes. The red ellipses represent the predicted standard deviation errors from a shot-noise-limited measurement of 7500 photons, as calculated from the Fisher information matrix. On average, the estimation errors for CD and sample orientation are 0.78 nm and 0.39$^\circ$, respectively. In general, the measurement error is expected to scale in proportion to the wavelength of illumination; this suggests, for example, that the average error for CD could be reduced to 0.39 nm by repeating the experiment with a green (532 nm) laser. This is on par with current industry needs, which demand measurements with accuracies on the order of a few angstroms. 

Notice also from Fig.~\ref{fig:param_retr} that the estimation errors for structures with similar true parameter values are highly correlated. In some cases (for example, the structure with 161 nm CD), this could signify inaccuracies in the assumed ``true'' parameter values and/or errors in other properties of the structure, such as the grating depth. Another likely contributor is the presence of systematic error (e.g., stress birefringence in the objective) that cannot be fully accounted for by the calibration procedure, which is solely based on measurements of the output intensity. Nevertheless, the relative errors between the two parameters (i.e., the error bar orientations) exhibit similar behavior to the Poisson statistical model. Comparing to Fig.~\ref{fig:in_out_intensities}(b), one can see that the estimation error is generally smallest when the output intensity is lowest, which occurs for sample orientations near $+6^\circ$. Again, this is consistent with the statistical model, which predicts small errors under low-light conditions due to the large fractional change in intensity associated with parameter variations. Note that the specific variations in intensity over the parameter space observed in this measurement are not a fundamental feature of the measurement scheme, but rather a consequence of the geometry of the sample and the chosen input polarization and analyzer. It is possible to define polarization distributions that, with more versatile polarization control, enable even more accurate parameter estimates, as demonstrated below.

\begin{figure}
	\centering
	\includegraphics[width=.9\linewidth]{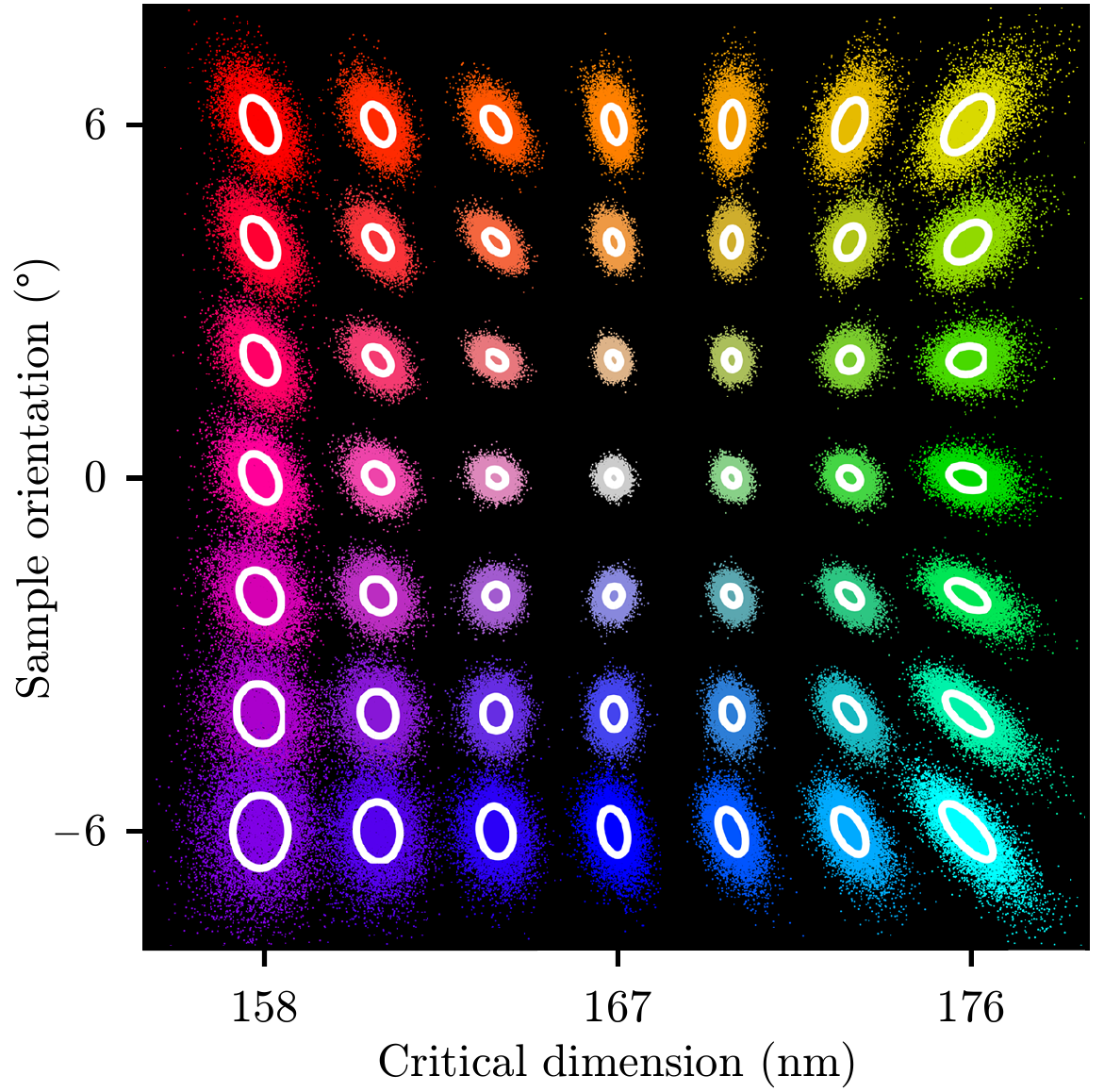}
	\caption{Retrieved parameters from output intensities containing 1000 photons, simulated for the input polarization and analyzer shown in Fig.~\ref{fig:in_out_intensities}(d). Data is shown for 10000 trials over 49 true parameter values (each shown in a different color). The ellipses represent the expected standard deviation errors for each true parameter value.}
	\label{fig:sim_param_retr}
\end{figure}

In order to predict the accuracy of future experiments using the optimized elliptical analyzer and spatially-varying input polarization shown in Fig.~\ref{fig:in_out_intensities}(d), we performed a Monte Carlo simulation in which the structure parameters were estimated from simulated intensity distributions containing a discrete number of photons. The results for 1000 photons are shown in Fig.~\ref{fig:sim_param_retr}, along with ellipses representing the expected standard deviation error. 
By repeating the simulation for 7500 photons, the performance of the optimal solution can be compared against the approximate error of the current experimental implementation. The most dramatic improvement occurs for the nominal structure having 167 nm CD and 0$^\circ$ orientation; for this case, the experimental standard deviation confidence intervals (based on the calibrated intensity profile) are $\pm 0.70$ nm and $\pm 0.35^\circ$. Under optimal conditions, these intervals are reduced to $\pm 0.06$ nm and $\pm 0.04^\circ$. For all values within the parameter ranges of interest, the error in CD is reduced by at least a factor of 3, while the orientation error is reduced by at least a factor of 1.25. Additional simulations corroborate that over smaller parameter ranges (for example, $\pm 3$ nm CD and $\pm 1^\circ$ rotation), an optimized spatially-varying input polarization could provide an even more significant advantage over a spatially uniform one. As mentioned earlier, the reason for this improvement is that a spatially-varying polarization can be optimized at each point to produce an output intensity with a larger fractional change with respect to variations in each parameter.

In summary, we have described a weak-measurement-inspired technique for the simultaneous measurement of multiple parameters. The specific implementation of this technique was a focused beam scatterometry experiment in which preselection and postselection were achieved via polarization control. Initial experiments involving a grating with 0.4 $\mu$m period demonstrate that even with simplistic polarization control, this method can produce measurements of physical parameter variations with subnanometer precision. This compares favorably with existing coherent Fourier scatterometry techniques, which have been used to perform similar measurements with uncertainties of one to two nanometers \cite{Kumar_2014}. Recent advances in the semiconductor industry have enabled the production of structures with periods of 20 nm and below \cite{ITRS_2015}; on this smaller scale, the scatterometry measurement presented here is expected to provide even greater sensitivity since a given physical variation would represent a larger relative change in the structure geometry. Future experiments with improved polarization control and/or shorter illumination wavelengths are also expected to further improve the accuracy of the measurement. These measurements may include additional parameters such as grating depth and sidewall angle, which will test the viability of the method for the several-parameter case. 

\begin{acknowledgments}
This work was carried out under a joint services agreement with IBM Corporation.  Supplemental funding was provided by New York State (NYSTAR) through the Center for Emerging and Innovative Systems and by the National Science Foundation (PHY-1068325, PHY-1507278). MAA received funding from the Excellence Initiative of Aix-Marseille University - A$^*$MIDEX, a French ``Investissements d'Avenir'' programme.

The authors would like to acknowledge Andrew Jordan and Philippe R\'efr\'egier for useful discussions and Jon Ellis, Steve Gillmer and Mike Theisen for their contributions to the experimental setup.
\end{acknowledgments}



\begin{thebibliography}{19}%
\makeatletter
\providecommand \@ifxundefined [1]{%
 \@ifx{#1\undefined}
}%
\providecommand \@ifnum [1]{%
 \ifnum #1\expandafter \@firstoftwo
 \else \expandafter \@secondoftwo
 \fi
}%
\providecommand \@ifx [1]{%
 \ifx #1\expandafter \@firstoftwo
 \else \expandafter \@secondoftwo
 \fi
}%
\providecommand \natexlab [1]{#1}%
\providecommand \enquote  [1]{``#1''}%
\providecommand \bibnamefont  [1]{#1}%
\providecommand \bibfnamefont [1]{#1}%
\providecommand \citenamefont [1]{#1}%
\providecommand \href@noop [0]{\@secondoftwo}%
\providecommand \href [0]{\begingroup \@sanitize@url \@href}%
\providecommand \@href[1]{\@@startlink{#1}\@@href}%
\providecommand \@@href[1]{\endgroup#1\@@endlink}%
\providecommand \@sanitize@url [0]{\catcode `\\12\catcode `\$12\catcode
  `\&12\catcode `\#12\catcode `\^12\catcode `\_12\catcode `\%12\relax}%
\providecommand \@@startlink[1]{}%
\providecommand \@@endlink[0]{}%
\providecommand \url  [0]{\begingroup\@sanitize@url \@url }%
\providecommand \@url [1]{\endgroup\@href {#1}{\urlprefix }}%
\providecommand \urlprefix  [0]{URL }%
\providecommand \Eprint [0]{\href }%
\providecommand \doibase [0]{http://dx.doi.org/}%
\providecommand \selectlanguage [0]{\@gobble}%
\providecommand \bibinfo  [0]{\@secondoftwo}%
\providecommand \bibfield  [0]{\@secondoftwo}%
\providecommand \translation [1]{[#1]}%
\providecommand \BibitemOpen [0]{}%
\providecommand \bibitemStop [0]{}%
\providecommand \bibitemNoStop [0]{.\EOS\space}%
\providecommand \EOS [0]{\spacefactor3000\relax}%
\providecommand \BibitemShut  [1]{\csname bibitem#1\endcsname}%
\let\auto@bib@innerbib\@empty
\bibitem [{\citenamefont {Aharonov}\ \emph {et~al.}(1988)\citenamefont
  {Aharonov}, \citenamefont {Albert},\ and\ \citenamefont
  {Vaidman}}]{Aharonov_1988}%
  \BibitemOpen
  \bibfield  {author} {\bibinfo {author} {\bibfnamefont {Y.}~\bibnamefont
  {Aharonov}}, \bibinfo {author} {\bibfnamefont {D.~Z.}\ \bibnamefont
  {Albert}}, \ and\ \bibinfo {author} {\bibfnamefont {L.}~\bibnamefont
  {Vaidman}},\ }\href@noop {} {\bibfield  {journal} {\bibinfo  {journal}
  {Phys.~Rev.~Lett.}\ }\textbf {\bibinfo {volume} {60}},\ \bibinfo {pages}
  {1351} (\bibinfo {year} {1988})}\BibitemShut {NoStop}%
\bibitem [{\citenamefont {Tamir}\ and\ \citenamefont
  {Cohen}(2013)}]{Tamir_2013}%
  \BibitemOpen
  \bibfield  {author} {\bibinfo {author} {\bibfnamefont {B.}~\bibnamefont
  {Tamir}}\ and\ \bibinfo {author} {\bibfnamefont {E.}~\bibnamefont {Cohen}},\
  }\href@noop {} {\bibfield  {journal} {\bibinfo  {journal} {Quanta}\ }\textbf
  {\bibinfo {volume} {2}},\ \bibinfo {pages} {7} (\bibinfo {year}
  {2013})}\BibitemShut {NoStop}%
\bibitem [{\citenamefont {Svensson}(2013)}]{Svensson_2013}%
  \BibitemOpen
  \bibfield  {author} {\bibinfo {author} {\bibfnamefont {B.~E.}\ \bibnamefont
  {Svensson}},\ }\href@noop {} {\bibfield  {journal} {\bibinfo  {journal}
  {Quanta}\ }\textbf {\bibinfo {volume} {2}},\ \bibinfo {pages} {18} (\bibinfo
  {year} {2013})}\BibitemShut {NoStop}%
\bibitem [{\citenamefont {Hosten}\ and\ \citenamefont
  {Kwiat}(2008)}]{Hosten_2008}%
  \BibitemOpen
  \bibfield  {author} {\bibinfo {author} {\bibfnamefont {O.}~\bibnamefont
  {Hosten}}\ and\ \bibinfo {author} {\bibfnamefont {P.}~\bibnamefont {Kwiat}},\
  }\href@noop {} {\bibfield  {journal} {\bibinfo  {journal} {Science}\ }\textbf
  {\bibinfo {volume} {319}},\ \bibinfo {pages} {787} (\bibinfo {year}
  {2008})}\BibitemShut {NoStop}%
\bibitem [{\citenamefont {Dixon}\ \emph {et~al.}(2009)\citenamefont {Dixon},
  \citenamefont {Starling}, \citenamefont {Jordan},\ and\ \citenamefont
  {Howell}}]{Dixon_2009}%
  \BibitemOpen
  \bibfield  {author} {\bibinfo {author} {\bibfnamefont {P.~B.}\ \bibnamefont
  {Dixon}}, \bibinfo {author} {\bibfnamefont {D.~J.}\ \bibnamefont {Starling}},
  \bibinfo {author} {\bibfnamefont {A.~N.}\ \bibnamefont {Jordan}}, \ and\
  \bibinfo {author} {\bibfnamefont {J.~C.}\ \bibnamefont {Howell}},\
  }\href@noop {} {\bibfield  {journal} {\bibinfo  {journal} {Phys.~Rev.~Lett.}\
  }\textbf {\bibinfo {volume} {102}},\ \bibinfo {pages} {173601} (\bibinfo
  {year} {2009})}\BibitemShut {NoStop}%
\bibitem [{\citenamefont {Dennis}\ and\ \citenamefont
  {G{\"o}tte}(2012)}]{Dennis_2012}%
  \BibitemOpen
  \bibfield  {author} {\bibinfo {author} {\bibfnamefont {M.~R.}\ \bibnamefont
  {Dennis}}\ and\ \bibinfo {author} {\bibfnamefont {J.~B.}\ \bibnamefont
  {G{\"o}tte}},\ }\href@noop {} {\bibfield  {journal} {\bibinfo  {journal} {New
  J. Phys.}\ }\textbf {\bibinfo {volume} {14}},\ \bibinfo {pages} {073013}
  (\bibinfo {year} {2012})}\BibitemShut {NoStop}%
\bibitem [{\citenamefont {Jayaswal}\ \emph {et~al.}(2014)\citenamefont
  {Jayaswal}, \citenamefont {Mistura},\ and\ \citenamefont
  {Merano}}]{Merano_2014}%
  \BibitemOpen
  \bibfield  {author} {\bibinfo {author} {\bibfnamefont {G.}~\bibnamefont
  {Jayaswal}}, \bibinfo {author} {\bibfnamefont {G.}~\bibnamefont {Mistura}}, \
  and\ \bibinfo {author} {\bibfnamefont {M.}~\bibnamefont {Merano}},\ }\href
  {\doibase 10.1364/OL.39.006257} {\bibfield  {journal} {\bibinfo  {journal}
  {Opt. Lett.}\ }\textbf {\bibinfo {volume} {39}},\ \bibinfo {pages} {6257}
  (\bibinfo {year} {2014})}\BibitemShut {NoStop}%
\bibitem [{\citenamefont {Hallaji}\ \emph {et~al.}(2016)\citenamefont
  {Hallaji}, \citenamefont {Feizpour}, \citenamefont {Dmochowski},
  \citenamefont {Sinclair},\ and\ \citenamefont {Steinberg}}]{Hallaji_2016}%
  \BibitemOpen
  \bibfield  {author} {\bibinfo {author} {\bibfnamefont {M.}~\bibnamefont
  {Hallaji}}, \bibinfo {author} {\bibfnamefont {A.}~\bibnamefont {Feizpour}},
  \bibinfo {author} {\bibfnamefont {G.}~\bibnamefont {Dmochowski}}, \bibinfo
  {author} {\bibfnamefont {J.}~\bibnamefont {Sinclair}}, \ and\ \bibinfo
  {author} {\bibfnamefont {A.~M.}\ \bibnamefont {Steinberg}},\ }\href@noop {}
  {\bibfield  {journal} {\bibinfo  {journal} {arXiv preprint arXiv:1612.04920}\
  } (\bibinfo {year} {2016})}\BibitemShut {NoStop}%
\bibitem [{\citenamefont {Zernike}(1942{\natexlab{a}})}]{Zernike_1942_1}%
  \BibitemOpen
  \bibfield  {author} {\bibinfo {author} {\bibfnamefont {F.}~\bibnamefont
  {Zernike}},\ }\href@noop {} {\bibfield  {journal} {\bibinfo  {journal}
  {Physica}\ }\textbf {\bibinfo {volume} {9}},\ \bibinfo {pages} {686}
  (\bibinfo {year} {1942}{\natexlab{a}})}\BibitemShut {NoStop}%
\bibitem [{\citenamefont {Zernike}(1942{\natexlab{b}})}]{Zernike_1942_2}%
  \BibitemOpen
  \bibfield  {author} {\bibinfo {author} {\bibfnamefont {F.}~\bibnamefont
  {Zernike}},\ }\href@noop {} {\bibfield  {journal} {\bibinfo  {journal}
  {Physica}\ }\textbf {\bibinfo {volume} {9}},\ \bibinfo {pages} {974}
  (\bibinfo {year} {1942}{\natexlab{b}})}\BibitemShut {NoStop}%
\bibitem [{\citenamefont {Murphy}(2001)}]{Murphy_2001}%
  \BibitemOpen
  \bibfield  {author} {\bibinfo {author} {\bibfnamefont {D.}~\bibnamefont
  {Murphy}},\ }\href@noop {} {\emph {\bibinfo {title} {Fundamentals of Light
  Microscopy and Digital Imaging}}}\ (\bibinfo  {publisher} {Wiley-Liss},\
  \bibinfo {address} {New York},\ \bibinfo {year} {2001})\ Chap.~\bibinfo
  {chapter} {10}, pp.\ \bibinfo {pages} {153--168}\BibitemShut {NoStop}%
\bibitem [{\citenamefont {Pedersen}\ and\ \citenamefont
  {Keller}(1986)}]{Pedersen_1986}%
  \BibitemOpen
  \bibfield  {author} {\bibinfo {author} {\bibfnamefont {K.}~\bibnamefont
  {Pedersen}}\ and\ \bibinfo {author} {\bibfnamefont {O.}~\bibnamefont
  {Keller}},\ }\href@noop {} {\bibfield  {journal} {\bibinfo  {journal} {Appl.
  Opt.}\ }\textbf {\bibinfo {volume} {25}},\ \bibinfo {pages} {226} (\bibinfo
  {year} {1986})}\BibitemShut {NoStop}%
\bibitem [{\citenamefont {Diebold}(2001)}]{Diebold_2001}%
  \BibitemOpen
  \bibfield  {author} {\bibinfo {author} {\bibfnamefont {A.~C.}\ \bibnamefont
  {Diebold}},\ }\href@noop {} {\emph {\bibinfo {title} {Handbook of silicon
  semiconductor metrology}}}\ (\bibinfo  {publisher} {CRC Press},\ \bibinfo
  {year} {2001})\BibitemShut {NoStop}%
\bibitem [{\citenamefont {Wilson}(2015)}]{ITRS_2015}%
  \BibitemOpen
  \bibfield  {author} {\bibinfo {author} {\bibfnamefont {L.}~\bibnamefont
  {Wilson}},\ }\href@noop {} {\emph {\bibinfo {title} {International Technology
  Roadmap for Semiconductors}}}\ (\bibinfo  {publisher} {Semiconductor Industry
  Association},\ \bibinfo {year} {2015})\BibitemShut {NoStop}%
\bibitem [{\citenamefont {den Boef}(2016)}]{DenBoef_2016}%
  \BibitemOpen
  \bibfield  {author} {\bibinfo {author} {\bibfnamefont {A.~J.}\ \bibnamefont
  {den Boef}},\ }\href@noop {} {\bibfield  {journal} {\bibinfo  {journal}
  {Surf. Topogr.: Metrol. Prop.}\ }\textbf {\bibinfo {volume} {4}},\ \bibinfo
  {pages} {023001} (\bibinfo {year} {2016})}\BibitemShut {NoStop}%
\bibitem [{\citenamefont {Boher}\ \emph {et~al.}(2004)\citenamefont {Boher},
  \citenamefont {Luet}, \citenamefont {Leroux}, \citenamefont {Petit},
  \citenamefont {Barritault}, \citenamefont {Hazart},\ and\ \citenamefont
  {Chaton}}]{Boher_2004}%
  \BibitemOpen
  \bibfield  {author} {\bibinfo {author} {\bibfnamefont {P.}~\bibnamefont
  {Boher}}, \bibinfo {author} {\bibfnamefont {M.}~\bibnamefont {Luet}},
  \bibinfo {author} {\bibfnamefont {T.}~\bibnamefont {Leroux}}, \bibinfo
  {author} {\bibfnamefont {J.}~\bibnamefont {Petit}}, \bibinfo {author}
  {\bibfnamefont {P.}~\bibnamefont {Barritault}}, \bibinfo {author}
  {\bibfnamefont {J.}~\bibnamefont {Hazart}}, \ and\ \bibinfo {author}
  {\bibfnamefont {P.}~\bibnamefont {Chaton}},\ }in\ \href@noop {} {\emph
  {\bibinfo {booktitle} {Metrology, Inspection, and Process Control for
  Microlithography XVIII}}},\ Vol.\ \bibinfo {volume} {5375}\ (\bibinfo
  {organization} {International Society for Optics and Photonics},\ \bibinfo
  {year} {2004})\ pp.\ \bibinfo {pages} {1302--1314}\BibitemShut {NoStop}%
\bibitem [{\citenamefont {Kumar}\ \emph {et~al.}(2014)\citenamefont {Kumar},
  \citenamefont {Petrik}, \citenamefont {Ramanandan}, \citenamefont
  {El~Gawhary}, \citenamefont {Roy}, \citenamefont {Pereira}, \citenamefont
  {Coene},\ and\ \citenamefont {Urbach}}]{Kumar_2014}%
  \BibitemOpen
  \bibfield  {author} {\bibinfo {author} {\bibfnamefont {N.}~\bibnamefont
  {Kumar}}, \bibinfo {author} {\bibfnamefont {P.}~\bibnamefont {Petrik}},
  \bibinfo {author} {\bibfnamefont {G.~K.}\ \bibnamefont {Ramanandan}},
  \bibinfo {author} {\bibfnamefont {O.}~\bibnamefont {El~Gawhary}}, \bibinfo
  {author} {\bibfnamefont {S.}~\bibnamefont {Roy}}, \bibinfo {author}
  {\bibfnamefont {S.~F.}\ \bibnamefont {Pereira}}, \bibinfo {author}
  {\bibfnamefont {W.~M.}\ \bibnamefont {Coene}}, \ and\ \bibinfo {author}
  {\bibfnamefont {H.~P.}\ \bibnamefont {Urbach}},\ }\href@noop {} {\bibfield
  {journal} {\bibinfo  {journal} {Opt. Express}\ }\textbf {\bibinfo {volume}
  {22}},\ \bibinfo {pages} {24678} (\bibinfo {year} {2014})}\BibitemShut
  {NoStop}%
\bibitem [{\citenamefont {Vella}()}]{Vella_2018_MLE_arxiv}%
  \BibitemOpen
  \bibfield  {author} {\bibinfo {author} {\bibfnamefont {A.}~\bibnamefont
  {Vella}},\ }\href@noop {} {\bibinfo  {journal} {arXiv:1806.04503}\
  }\BibitemShut {NoStop}%
\end{thebibliography}

%

\FloatBarrier
\clearpage

\onecolumngrid
\begin{center}\Large \bf Appendix\end{center}
\twocolumngrid

\section{I.\quad Experimental details}\label{app_1}
\begin{figure}[h]
	\centering
	\includegraphics[width=\linewidth]{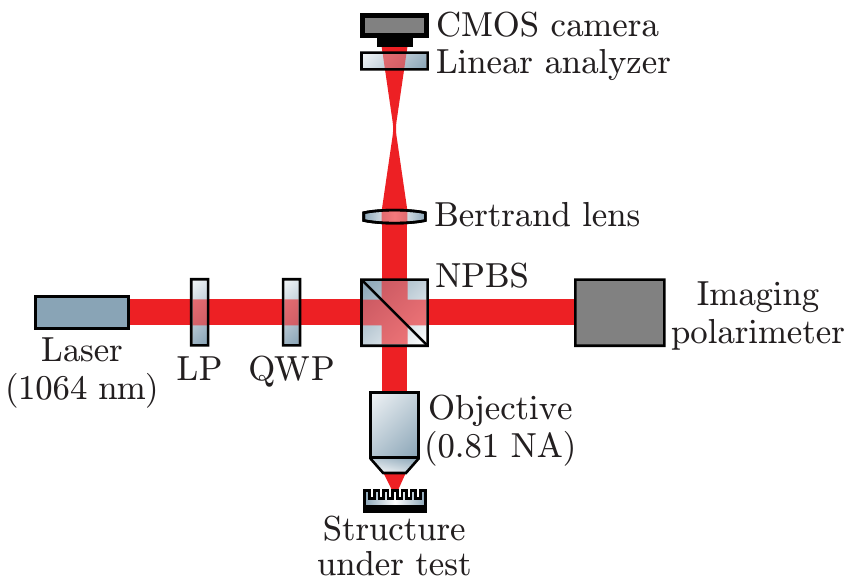}
	\caption{Schematic of the experimental setup used for a two-parameter measurement of a silicon lamellar grating. LP = linear polarizer, QWP = quarter-wave plate, NPBS = non-polarizing beamsplitter.}
	\label{fig:FBS_schematic_expt}
\end{figure}

A more detailed schematic of the experimental setup is shown in Fig.~\ref{fig:FBS_schematic_expt}. The polarization generator consists of a linear polarizer and quarter-wave plate, which may be rotated to generate any spatially uniform elliptical polarization state. The resulting polarization was measured (after transmission through the non-polarizing beamsplitter) using an imaging polarimeter consisting of a rotating quarter-wave plate and a fixed linear polarizer. A Bertrand lens was used to image the pupil of the objective onto the detector.

\phantom{The optimized functions $\pbar_1(\bu)$ and $\pbar_2(\bu)$ associated with the input polarization shown in Fig.~2(d) in the main text are plotted in Fig.~\ref{fig:pbars} below. These functions can be interpreted as the departure from perfect nulling (i.e., zero output intensity) associated with each parameter, normalized to the range of interest. In other words, they}

\FloatBarrier
\newpage
\FloatBarrier

\section{II.\quad Optimized pupil functions}\label{app_2}
The optimized functions $\pbar_1(\bu)$ and $\pbar_2(\bu)$ associated with the input polarization shown in Fig.~2(d) in the main text are plotted in Fig.~\ref{fig:pbars} below. These functions can be interpreted as the departure from perfect nulling (i.e., zero output intensity) associated with each parameter, normalized to the range of interest. In other words, they determine the range of variation associated with each weak value of $\R_n$ at each point in the pupil. The most important feature of these plots is that each parameter has a pupil function with distinct spatial variations on the order of unity, enabling accurate estimates over the entire range of interest with minimum coupling between parameters.

\begin{figure}[h]
	\centering
	\includegraphics[width=.98\linewidth]{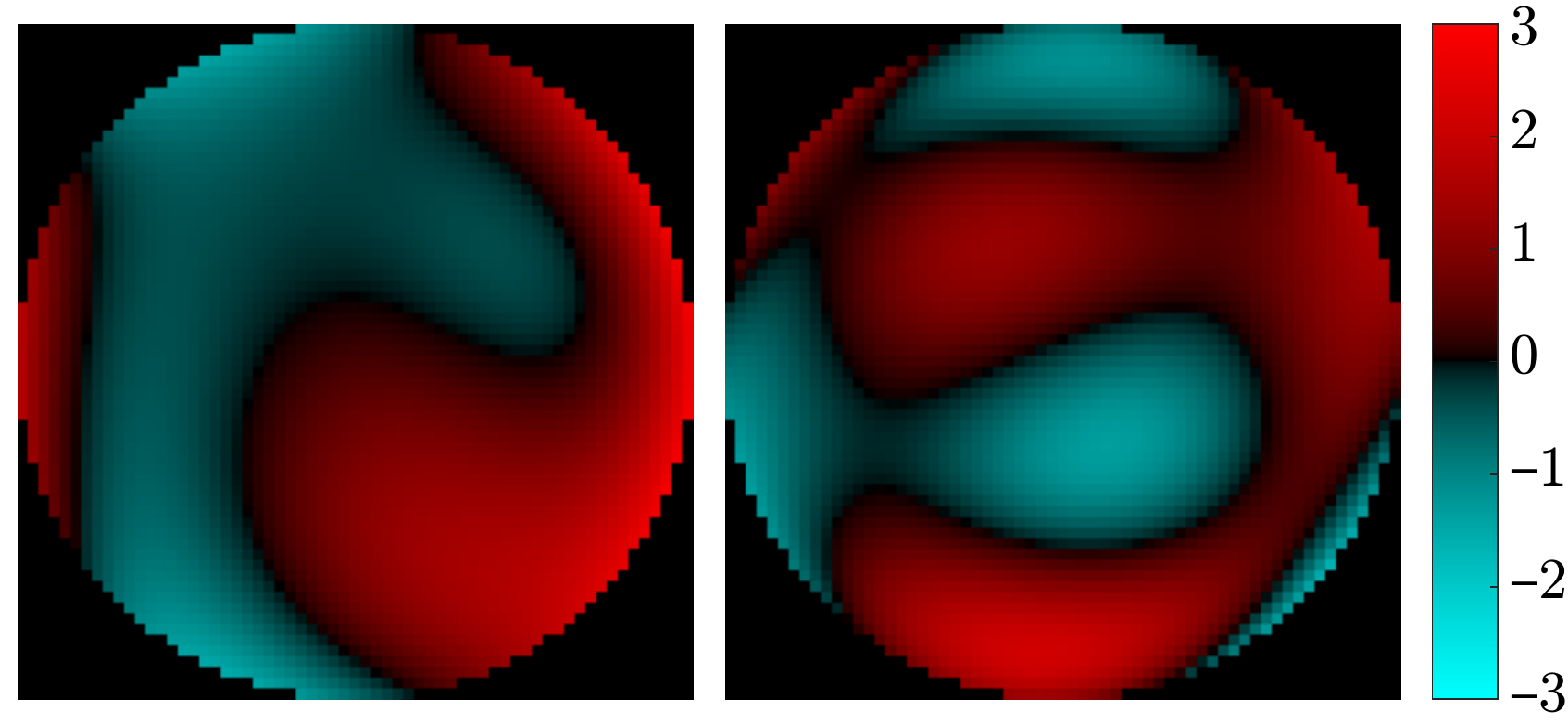}
	\caption{Optimized pupil functions $\pbar_1(\bu)$ (left) and $\pbar_2(\bu)$ (right) for a measurement of CD and sample orientation using the elliptical analyzer shown in the inset of Fig.~2(d) in the main text.}
	\label{fig:pbars}
\end{figure}

\end{document}